# Three-dimensional atom-by-atom mapping of nanoscale precipitates in single Te inclusions in $Cd_{0.9}Zn_{0.1}Te$ crystal

Eloïse Rahier, Sebastian Koelling, Guillaume Nadal, Sudarshan Singh, Luc Montpetit, Oussama Moutanabbir

*Department of Engineering Physics, Polytechnique Montreal, C. P. 6079, Succ. Centre-Ville, Montreal, QC H3C 3A7, Canada*

Abstract:
The complexity and richness of phenomena governing alloy crystal growth can be unraveled by examining the three-dimensional atomic-level distribution of elements and impurities incorporated during growth. These species act as atomic fingerprints, revealing the thermodynamic constraints that shape material structure and composition. Herein, we combine transmission electron microscopy and atom probe of tellurium (Te) inclusions within cadmium zinc telluride (CZT) single crystals. The correlative analysis uncovers nanoscale precipitates embedded within Te inclusions, consisting of CZT nanocrystals with a Zn content of 1.5 at.%. Surrounding these precipitates, a ~10 nm-thick shell is observed, enriched with copper and indium impurities. In addition, traces of sodium and sulfur are detected within the nanocrystals. These findings provide direct evidence of the complex segregation and precipitation processes occurring during CZT crystal growth, reflecting the interplay of thermodynamic driving forces and kinetic constraints that govern solute redistribution. The resulting insights contribute to a deeper understanding of impurity behavior and phase separation mechanisms in CZT alloys. This work establishes a framework for modeling and optimization of growth strategies of higher-quality CZT crystals for next-generation infrared and radiation detection technologies.

Cadmium Zinc Telluride (CZT) has emerged as a leading compound semiconductor for radiation detection and optoelectronic applications, owing to its unique combination of relatively wide bandgap, high atomic number constituents, and favorable electronic transport properties.[1,2] Unlike traditional X-ray detector materials that require cryogenic cooling, CZT supports room-temperature operation while delivering excellent energy resolution, making it particularly attractive for compact and field-deployable devices. One of the primary domains where CZT has demonstrated significant impact is in radiation detection, including spectroscopy and imaging.[1] Its high resistivity and efficient photon absorption enable precise energy discrimination, which is essential for applications such as medical imaging (*e.g.*, SPECT and CT scanners),[3,4] nuclear security,[5] and environmental monitoring.[1,6] In the field of security, CZT-based detectors are deployed in portable spectrometers and radiation portal monitors for the identification of illicit radioactive sources, offering both spectral resolution and operational robustness.[7,8]

In photonics and optoelectronics, CZT also serves as a lattice-matched substrate for the epitaxial growth of HgCdTe, a material widely used in mid-infrared photodetectors.[9,10] This makes CZT foundational for the fabrication of advanced infrared focal plane arrays used in thermal imaging, night vision, guidance systems, and several optoelectronic platforms.[10-12] Beyond terrestrial applications, CZT plays a critical role in astrophysics and space science. Its deployment in spaceborne telescopes[13,14] has enabled high-resolution imaging and spectroscopy of cosmic X-ray and gamma-ray sources, facilitating studies in high-energy astrophysical phenomena and cosmology. Given this broad technological relevance and the growing demand for high-performance, room-temperature radiation detectors and substrates, the continued development of CZT, including improvements in crystal quality, compositional uniformity, and defect control, remains a key priority.

Developing CZT devices at scale demands crystal growth methods capable of producing large volumes of high-quality materials. The most used techniques today are the Bridgman methods (both conventional and modified)[10,15,16] and the Travelling Heater Method (THM).[10,17] While the Bridgman techniques offer faster growth rates, the high temperatures involved can introduce contaminants from the crucible and growth environment.[18,19] In contrast, THM yields the highest material quality but at a higher production cost.[18] Nevertheless, both methods commonly result in residual defects, notably Te-rich secondary phases, commonly known as Te inclusions, which can impact the device behavior and performance. In fact, Te inclusions may impair optical transmission, electrical recombination, mobility-lifetime product and more.[20,21] Early studies focus on how growth conditions influence the spatial distribution of these defects to ultimately eliminate them.[16,22] The general consensus is that these inclusions contain impurities,[19,23] however, little is known about their three-dimensional atomic-scale distribution. These details are essential to track the thermodynamic constraints limiting the uniformity and purity of CZT crystals.

Herein, Transmission Electron Microscopy (TEM) and Atom Probe Tomography (APT) analyses are combined to achieve unprecedented three-dimensional atomic-scale mapping of Te inclusions within the CZT lattice. These atomic-level investigations reveal the existence of nanoscale precipitates within these inclusions. The precipitates consist of CdZnTe nanocrystals with a Zn content of 1.5 at.% and a 10 nm-thin shell containing Cu and In impurities. Traces of Na and S are also detected in Te inclusions. The resulting insights are critical to elucidate the mechanisms underlying the crystallization process that will ultimately guide the modeling and optimization of growth methods toward higher-quality CZT crystals.

CZT ($Cd_{0.9}Zn_{0.1}Te$) single crystals used in this study were grown by THM. To get a large-scale distribution of the inclusions, infrared images of the investigated samples were first acquired in transmission mode using an IR microscope. It was equipped with a halogen bulb and an infrared camera. Microstructural investigations and atomic-level chemical analysis were carried out using TEM and APT. APT allows the three-dimensional reconstruction of tip-shaped specimens by sequentially removing individual atoms through field evaporation assisted by a pulsed laser. Each evaporated atom is chemically identified and spatially reconstructed based on its time-of-flight and impact position on the detector, providing atomic-scale compositional and structural information.[24] APT measurements were performed using an Invizo 6000 system equipped with a picosecond laser at a wavelength of 257.5nm and a double einzel lens to focus the ions on the micro-channel plate[25] detector allowing for a wide field of view.[26] During the analysis, the samples were cooled down to a base temperature of 25 K. The experimental data were collected at a laser pulse rate of 200 kHz and a laser power of 1.5-12 pJ. The data are reconstructed into 3D representations of the analyzed materials in IVAS6.3 using a standard protocol.[27] TEM lamellas and APT tips were prepared using Focused Ion Beam (FIB) following the standard lift-out procedures and final milling at 2 kV. Inclusions of tens of micrometers in size were targeted. Conventional TEM images were acquired with a JEOL JEM-F200. High-Angle Annular Dark-Field Scanning Transmission Electron Microscopy (HAADF-STEM) and Energy-Dispersive X-ray Spectroscopy (EDS-STEM) analyses were performed using a Thermofisher Talos F200X G2. Both microscopes were operated with a 200 keV electron beam.

Infrared imaging (Figure 1(a)) reveals that the CZT matrix is decorated by Large Te inclusions appearing as dark spots exceeding several tens of micrometers in size. Figure 1(b) presents a scanning electron microscopy (SEM) image acquired during the preparation of a TEM lamella, with carbon and platinum layers deposited to protect the region of interest. A cross-sectional specimen was prepared across the two inclusions highlighted in yellow, with most of the section consisting of the surrounding CZT matrix. An overview of the final TEM lamella is shown in Figure 1(c), where the CZT crystal and the inclusion (bottom left) are distinguishable due to their different contrasts. Figure 1(d) and Figure 1(e) are Electron Diffraction (ED) patterns of the CZT matrix and the inclusion, respectively. The ED pattern is a direct image of the reciprocal space of a crystal lattice where the distance from the central spot (covered) to a diffraction spot is directly the inverse of the plane distance in the crystal (d-spacing). Both patterns exhibit single crystal characteristics and the measured d-spacings are consistent with the cubic zincblende structure of CZT (3.69 Å and 3.23 Å for {111} and {200} plane family respectively) and the hexagonal structure of Te (3.87 Å and 2.33 Å for {100} and {102} plane family respectively).

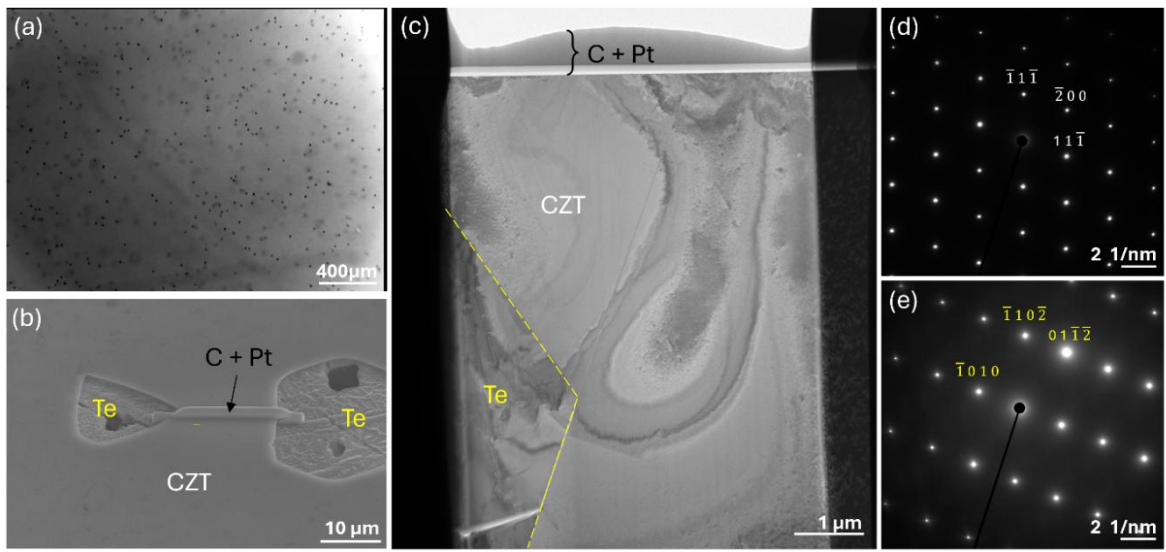

**Figure 1:** (a) Infrared image of a CZT wafer showing Te inclusions as darker spots. (b) SEM image of the lamella preparation for TEM characterization. The surface of the sample was covered with C and Pt to protect the sample during FIB operation. (c) Cross-sectional TEM image of a CZT sample in the vicinity of a Te inclusion. (d) ED pattern of a CZT area oriented along the [011] zone axis and (e) ED pattern of the Te inclusion oriented along the [$\bar{2}$ 4 $\bar{2}$ 3] zone axis. ED patterns are indexed to show the main three crystallographic directions observed.

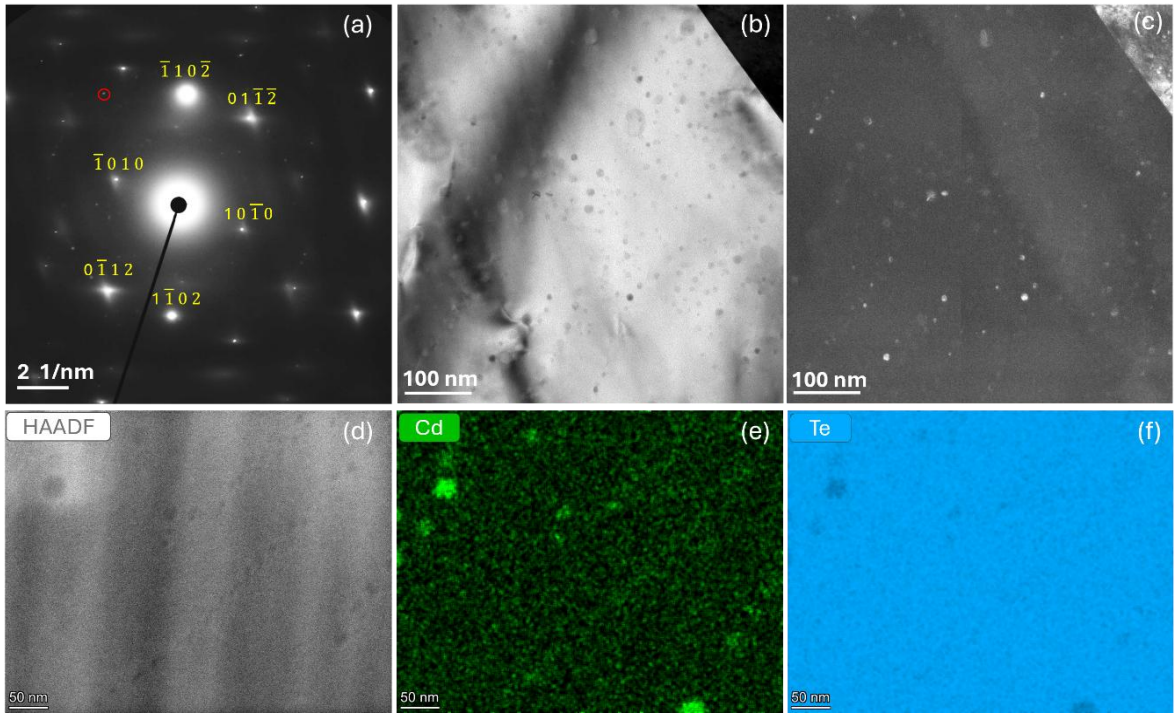

**Figure 2:** (a) ED pattern, (b) brightfield (BF) image, (c) darkfield (DF) image, (d) HAADF-STEM image, (e) EDS-STEM map of Cd and (f) Te, taken in the Te inclusion. DF was obtained by using the diffraction spot circled in the diffraction pattern.

Figures 2(a-c) show an ED pattern, brightfield (BF) and darkfield (DF) TEM images of the Te inclusion oriented close to the [$\bar{2}\,4\,\bar{2}\,3$] zone axis. In BF imaging, contrast arises from unscattered electrons. However, in DF imaging, contrast is generated by exclusively selecting electrons that have been scattered by the specimen under specific diffraction conditions, while excluding the unscattered electrons. The brighter spots indexed in yellow on the ED pattern (Figure 2(a)) originate from the diffraction by Te single crystal. Additional spots come from nanoscale precipitates visible as dark, round features in the BF image (Figure 2(b)). The DF image (Figure 2(c)) is obtained using the diffraction spot circled in red in the ED pattern, corresponding to a 1.95 Å d-spacing. The bright features in the DF image correspond to the precipitates observed in the BF image. These precipitates have sizes ranging from approximately 10 to 50 nm. The CZT matrix, visible in the upper-right corner, is also diffracting under the same conditions. The measured lattice spacing of 1.95 Å matches the {311} family

of planes for CZT, indicating that the precipitates within the Te inclusion may possess a composition close to that of CZT.

EDS analyses were performed in scanning mode to acquire two-dimensional elemental distribution maps. In parallel, a HAADF image was generated by collecting electrons scattered at high angles. The resulting image contrast is strongly dependent on the atomic number of the elements present. Figures 2(d-f) display a HAADF-STEM image alongside the corresponding EDS-STEM elemental maps obtained from within a single Te inclusion. In the HAADF image, precipitates appear as darker, circular regions, indicating the presence of elements with lower atomic numbers compared to the surrounding Te-rich matrix. The associated EDS-STEM maps (Figure 2(e) and Figure 2(f)) support this observation showing that these regions are depleted in Te and enriched in Cd. However, due to the limited spatial resolution and sensitivity of the EDS analysis, a precise chemical composition of those precipitates could not be determined. To address this limitation, the three-dimensional atomic-scale chemical composition of the precipitates were investigated by APT.

Figure 3(a) presents the SEM image of a representative APT specimen prior to the field evaporation. The tip was shaped from a 10 µm wide Te inclusion identical to the one displayed on the right side of Figure 1(b). The atom-by-atom reconstructions are presented in Figures 3(b-e) showing Cd, Zn and Te atoms together and individually. The specimen is mainly composed of Te and contains four Cd and Zn enriched clusters clearly discernible in the Cd and Zn-only reconstructions (Figures 3(d,e)). The high noise seen in the Zn-only reconstruction arises from overlap of peaks from different ions in the mass spectrum. These overlaps could not be considered for atomic representations. However, the contribution of each atomic species to every overlapping peak has been accounted for in the compositional profiles presented later.

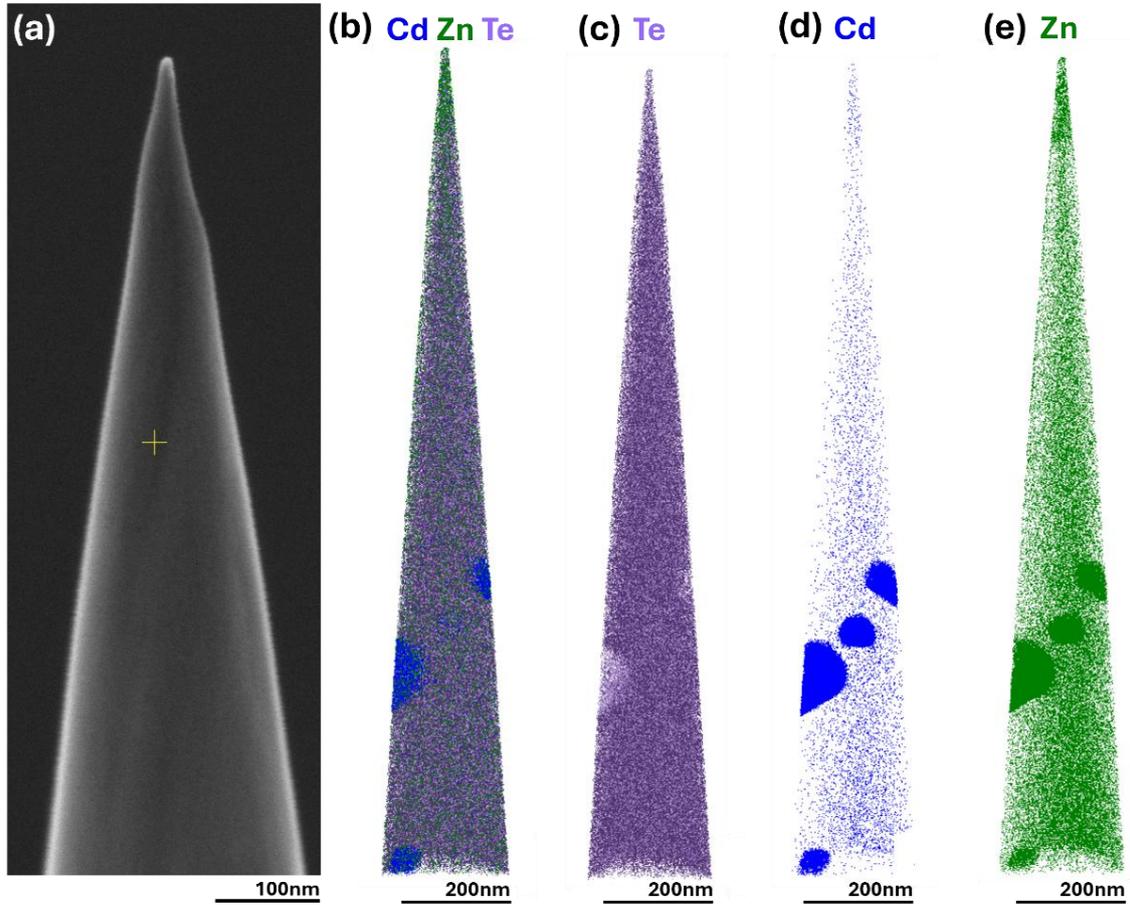

**Figure 3:** (a) SEM image of an APT specimen prepared using FIB from a single Te inclusion. (b) Three-dimensional atom-by-atom reconstructions of a typical CZT specimen showing Te, Cd, and Zn atoms together along with their individual maps separated (c-e). The analyzed volume consists of 266,946,388 atoms.

Figures 4(a-d) show a sub-volume reconstruction of this measurement centered around the three clusters in the middle of the APT tip and its associated mass spectrum (Figure 4(e)). Along with the expected CdZnTe components, four different impurities were identified: Cu, In, Na and S. The atom-by-atom reconstructions (Figure 4(a-c)) unravel the spatial position of these impurities. In and Cu atoms surround the Cd-rich precipitates. Two small Na clusters, comprising respectively about 30 and 500 atoms, are randomly distributed in the Te inclusion. One is comprised within the cluster (on the right side of the reconstruction) while the other is outside the clusters (left side of the reconstruction). A compositional profile taken across the three Cd-rich regions (following the dotted arrow indicated in Figures 4(a-c)) is presented in Fig. 4.e. Within the precipitates, the Te content decreases from 100% to around 50%, while Cd emerges as the second most abundant element reaching 48%.

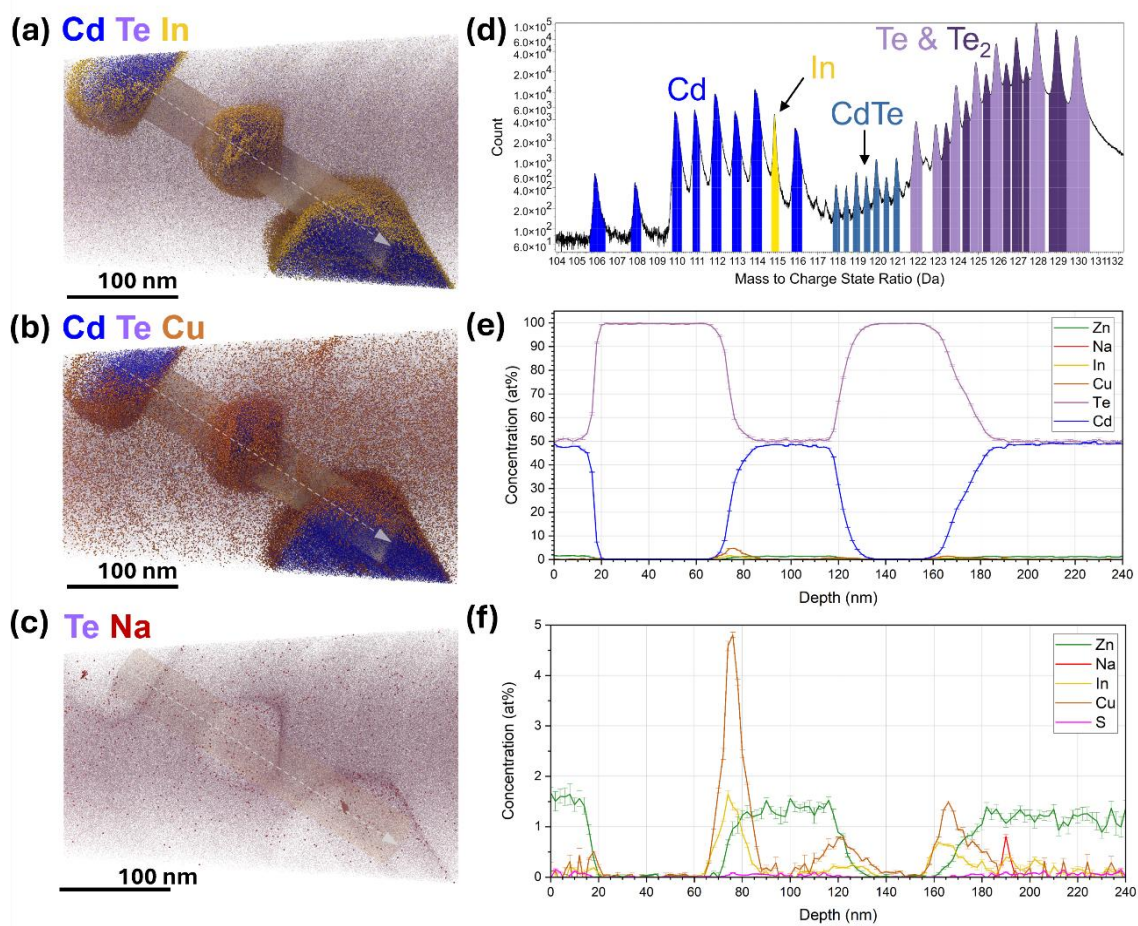

**Figure 4:** Atom-by-atom reconstructions of the sub-volume of the Te inclusion tip, centered around the three clusters, showing (a) Cd, Te and In, (b) Cd, Te and Cu and (c) Te and Na atoms. In, Cu, Na are colored yellow, dark orange and red respectively. (d) Mass spectrum of the sub-volume centered around the three clusters identifying Cd, In, CdTe and Te. (e) Concentration profile of Zn, Na, In, Cu, Te and Cd extracted from a 40 nm-diameter cylinder crossing the three clusters, materialized in the reconstructions (a-c). (f) Same concentration profiles in (e) highlighting the profiles of low-concentration elements.

Figure 4(f) highlights the profiles of low-concentration elements, confirming that the three clusters are in fact CZT nanocrystals with low Zn content of about 1.5 at.% . While In and Cu impurities are present in small quantities (below 0.1 at.%) within the CZT precipitates, their concentrations notably increase at the boundaries of said precipitates. The profile across one of the Na clusters peaks at a concentration of 1 at.% (Figure 4(f), red). The other cluster is outside the CZT precipitates and contains 30 atoms embedded within the Te inclusion. There are no detectable traces of Na in the rest of the measured volume. Traces of S (below 0.1 at%) are also detected inside the CZT crystals. However, S is homogeneously distributed within the clusters compared to Cu and In, which segregate on the surface. Notably, apart from the Na cluster discussed previously, no impurities were detected in the Te inclusion.

The THM growth process employs a Te-rich CZT molten zone in contact with a CZT seed to enable crystallization below the melting point of CZT (typically 900 °C). Operating at such reduced temperatures minimizes contamination from the growth environment[17] and promotes the formation of high-purity crystals. However, the growth–molten-zone interface remains a critical region where small perturbations can generate significant compositional inhomogeneities. As reported previously,[17,22] imperfections at the growth front or fluctuations in local composition can trap Te-rich liquid droplets within the solidifying CZT. Upon cooling, these droplets gradually solidify, forming the Te inclusions frequently observed in THM-grown ingots.

The detailed analyses described above reveal that these inclusions are not single-phase but exhibit a complex composite structure consisting mainly of CZT nanocrystals (~1.5 at.% Zn) embedded in a Te matrix and coated by a ~10 nm-thick shell enriched with In and Cu. As a first step, the sequence of phase evolution can be rationalized based on bulk CdTe phase diagram.[28] As the temperature decreases from 900 °C to 450 °C, CdTe crystals nucleate within the liquid Te, while excess Zn incorporates into their lattice, forming CZT nanocrystals. Upon further cooling, Te solidifies around these precipitates, encapsulating them within the crystalline lattice. Impurities already present in the polycrystalline chargeare, in principle, filtered by the Te-rich solvent zone. Because Te is a by-product of Cu mining, residual Cu remains present even after tellurium refining. In is intentionally introduced as a dopant in the polycrystalline charge. Na is a common contaminant associated with process handling, while S is known to originate from organic substances present during the various compounding steps of the CZT polycrystalline charge. The observed segregation of In and Cu at the CZT nanocrystal surface, alongside small Na clusters and homogeneously distributed S, reflects differences in solubility and surface energetics. Cu and In possess relatively low surface tensions, which favor their migration toward interfaces to minimize the overall free energy. Their accumulation at the CZT-Te interface is thus driven not only by their low solubility in solid CZT and Te but also by their intrinsic tendency to reduce interfacial energy. By contrast, Na and S, with smaller atomic radii, follow different segregation pathways: Na forms isolated clusters, while S is randomly distributed in CZT nanocrystals. During ingot growth, Te-rich droplets act as getters for impurities and excess Cd and Zn. The absence of detectable impurities within Te confirms that, upon cooling, Te expells these species, forming the observed composite structure.

It is also important to note that additional size-dependent thermodynamic effects can further modify these segregation phenomena. The CZT nanocrystals, averaging ~50 nm in diameter, possess a sufficiently high surface-to-volume ratio for surface energy to influence their phase stability. Surface tension contributes to a modest reduction in the local melting temperature and a shift in equilibrium phase boundaries relative to bulk CZT. The enhanced curvature amplifies the driving force for low-surface-tension elements such as In and Cu to segregate at the surface. These effects can promote compositional gradients and subtle core-shell-like architectures within individual nanocrystals. Altogether, surface energetics and size effects jointly govern the evolution and stabilization of Te inclusions, linking nanoscale thermodynamics to macroscopic crystal quality.

In summary, through a correlative analysis combining TEM and APT analyses of Te inclusions in CZT single crystals, we reveal nanoscale CZT precipitates with a Zn content of about 1.5 at.% embedded within the inclusions. These precipitates are surrounded by a ~10 nm-thick shell enriched with Cu and In impurities, while traces of Na and S are also detected. The spatial organization of these species provides direct evidence of impurity segregation and precipitation mechanisms. They are governed by competing thermodynamic driving forces and kinetic constraints during CZT crystal growth. Te inclusions are a result of liquid Te-rich droplets trapped at the solvent/crystal interface that getter impurities, actively purifying the CZT crystal being grown. The insights gained from these atomic-level observations establish a comprehensive picture of impurity behavior and phase separation in CZT alloys. More broadly, they lay the groundwork for predictive modeling of the fabrication of high-purity CZT crystals for next-generation infrared and radiation detection technologies.

ACKNOWLEDGMENT

The authors thank Jean-Philippe Masse for help with TEM imaging. This work was funded by NSERC Canada, Canada Foundation for Innovation, and PRIMA Québec.